\documentclass[letterpaper,10pt,twocolumn]{article}   
\usepackage{osajnl2} 
\usepackage[draft]{hyperref} 

\begin{document}

\twocolumn[ 
\title{Quantum efficiency measurement of single photon detectors using photon pairs generated in optical fibers}

\author{Xiaoying Li*, Xiaoxin Ma, Limei Quan, Lei Yang, Liang Cui and Xueshi Guo}%

\address {College of Precision Instrument and
Opto-electronics Engineering, Tianjin University, \\Key Laboratory of Optoelectronics Information Science and Technology, Ministry of Education, Tianjin, 300072,
P. R. China}

\email{*Corresponding author: xiaoyingli@tju.edu.cn}

\begin{abstract}
Using the correlated signal and idler photon pairs generated in a dispersion shifted fiber by a pulsed pump, we measure the quantum efficiency of a InGaAs/InP avalanche photodiode-based single photon detector. Since the collection efficiency of photon pairs is a key parameter to correctly deduce the quantum efficiency, we carefully characterize the collection efficiency by studying correlation dependence of photon pairs upon the spectra of pump, signal and idler photons. This study allows us to obtain quantum efficiency of the single photon detector by using photon pairs with various kinds of bandwidths.
\end{abstract}


\maketitle 
 ] 

\section{Introduction}

Single photon detectors are finding various important scientific and technological applications. In particular, because 1550 nm telecom band is the most attractive wavelength from the viewpoint of fiber transmission, InGaAs/InP avalanche photodiode-based single photon detectors (SPDs) are useful for quantum communications~\cite{Gisin02} and quantum state measurement~\cite{Achi06}. Therefore a
precise determination of their quantum efficiency (QE) is essential. Conventional calibrations performed by comparison to a reference standard are complex and difficult. Whereas quantum correlated photon pairs can be used to realize absolute calibration of photon counting detectors without any ties to externally calibrated standards.
The photon pairs based method for SPD efficiency measurement was proposed by Klyshko~\cite{Klyshko80} and had been experimentally demonstrated by a number of groups~\cite{Brida00,Migdall04,Worsley09}. The absolute nature of the method originates from the quantum correlation of photon pairs, the detection of one photon indicates with certainty the existence of the other. The measurement uncertainty of QE highly depends on the accuracy of the estimated losses of photon pairs, including the transmission and collection efficiencies.

It has been experimentally proven that spontaneous parametric processes, including spontaneous parametric down conversion (SPDC) in $\chi ^{(2)}$ nonlinear crystals~\cite{Weinberg70,ou88b,Kwait95} and spontaneous four wave mixing (SFWM) in $\chi ^{(3)}$ based nonlinear materials~\cite{Fiorentino02,Alibart06,Fan05a}, are efficient methods to generate the correlated photon pairs. Photon pairs generated by SPDC and SFWM exhibit similar entanglement properties because of their fundamental similarity, both of the two kinds of parametric processes are constrained by energy and momentum conservations. Therefore, the tasks of quantum information processing fulfilled by utilizing photon pairs via SPDC~\cite{Zeilinger00,Sergienko03,Kok07} should in principle be carried out by using photon pairs via SFWM as well~\cite{Chen08,Fan08,Rarity09a-opex}. So far, most experimental demonstrations of the direct efficiency measurements were realized by using photons pairs generated in $\chi ^{(2)}$ crystals~\cite{Rarity87, Kwait94-ao,Brida00,Worsley09} pumped by continuous wave lasers, and a high accuracy verification with a relative standard uncertainty of  $0.18\%$ had been recently demonstrated~\cite{Migdall07-opex}. However, the photon pairs generated in fibers via SFWM by pulsed lasers, having the advantage of modal purity~\cite{Alibart06, Fiorentino02}, have not been used to measure QE of SPDs. Moreover, the majority of absolute calibrations concerns the silicon avalanche photodiode-based SPDs, the direct efficiency measurement of InGaAs/InP avalanche photodiode-based SPD has yet to be fully experimentally exploited~\cite{Odate07}.

In this paper, using the correlated signal and idler photon pairs in 1550 nm band, which are generated in 300 m dispersion shifted fiber (DSF) by a pulsed pump, we perform the QE measurement of a InGaAs/InP avalanche photodiode-based SPD. Because of
the broadband nature of pulsed pump and bandwidth of SFWM in DSF, for the photons function as heralding photons, say idler photons with a certain narrow bandwidth, the spectrum of its corresponding signal photons has a broader bandwidth~\cite{Li08a-opex, Li06qcmc}. Therefore, to accurately evaluate collection efficiency of the photon pairs, we investigate the correlation dependence of photon pairs upon the spectra of pump, signal and idler photons. The investigation allows us to obtain quantum efficiency of the SPD by using photon pairs with various kinds of bandwidths. While on the contrary, for the QE measurements done by photon pairs via SPDC in $\chi ^{(2)}$ crystals~\cite{Rarity87, Kwait94-ao,Migdall07-opex}, the bandwidth of the heralded photons is required to be much broader than that of the heralding photons so that all the pair events of the heralding photons are caught. Since the QE measurement is based on a careful characterization of the spectra correlation of the photon pairs, our investigations are also useful for developing other photon pair based quantum information technologies.

The paper is organized as follows. After presenting the principle of the experiment in section 2, we briefly describe our photon counting system in section 3. In section 4, we will describe the experimental procedure and analyze and interpret the experimental data. In section 5, we analyze the uncertainties and reliability of the experimental results, and discuss the possibilities of the improvements. Finally, we briefly conclude in section 6.

\section{Experimental principle}

When the central wavelength of the pump pulses is in the anomalous-dispersion regime of DSF, phase-matching is satisfied and the probability of SFWM is significantly enhanced. In this process, two
pump photons at frequency $\omega _p$ scatter through the Kerr ($\chi ^{(3)}$)
nonlinearity of the fiber to create
energy-time entangled signal and idler photons at frequencies $\omega _s$ and $%
\omega _i$, respectively, such that $2\omega _p=\omega _s+\omega
_i$. Since the photons are created in pairs, the detection of one photon indicates the existence of the other. This fact allows one to make absolute determinations of detector quantum efficiency.

Because of the isotropic nature of the Kerr nonlinearity in
fused-silica-glass fiber, the generated photon pairs are
predominantly co-polarized with the pump photons. In the low gain regime, the two-photon state can be expressed as,
\begin{eqnarray}
\mid \Psi (\omega _s,\omega _i)\rangle \propto \int d\omega _s\int d\omega
_iF(\omega _s,\omega _i)\mid \omega _s\rangle \mid \omega _i\rangle ,
\label{state}
\end{eqnarray}
where $\mid \omega _s\rangle $ and $\mid \omega _i\rangle $ in Eq. (\ref{state}) are single-photon signal and idler states in which the photons are present at frequency $\omega _s$ and $\omega _i$. $F(\omega _s,\omega _i)$ is the two-photon spectral function, which is proportional to the probability amplitude of two-photon detection, can be written as
$F(\omega _s,\omega _i)= \alpha (\omega _s,\omega _i)\phi (\omega _s,\omega _i)$,
where $\alpha (\omega _s,\omega _i)$ is the pump field spectrum, and $\phi (\omega _s,\omega _i)$ describes the phase matching condition.

Signal and idler photons created in pairs can be detected by SPDs with QEs, $\eta_{d1} $ and $\eta_{d2} $, respectively. When pump is a single frequency continuous wave, frequencies of signal and idler are perfectly correlated. Similar to SPDC in $\chi ^{(2)}$ crystals~\cite{Rarity87},  for the signal and idler photons having identical bandwidths, when the pair production rate is $R_{FWM} $, the average detection rates in signal and idler channels, $N_s$ and $N_i$, can be written as
\begin{eqnarray}
N_s=\eta _{d1}\eta _{ts}R_{FWM},~~  N_i=\eta _{d2}\eta _{ti}R_{FWM},
\end{eqnarray}
where $\eta _{ts}$ and $\eta _{ti}$ are the transmission efficiencies in signal and idler channels, respectively. When $R_{FWM}$ is so small that the multi-pair events is  negligible, the mean coincidence is given by
\begin{eqnarray}
\overline{c}=\eta _{d1}\eta
_{ts}N_i=\eta _{d2}\eta _{ti}N_s.
\end{eqnarray}
Therefore, the QEs of SPD can be deduced from
\begin{eqnarray}
\eta _{d1}=\frac{\overline{c}}{N_i\eta _{ts}},~~\eta _{d2}=\frac{\overline{c}%
}{N_s\eta _{ti}}.
\label{effi-cw}
\end{eqnarray}
Eq. (\ref{effi-cw}) shows that when SPD in idler (signal) channel serves as a trigger detector to record the single counts $N_{i(s)}$, and the transmission efficiency $\eta _{ts}$ is correctly estimated, the efficiency of SPD in signal (idler) channel can be absolutely determined by a two-photon coincidence measurement.

In practice, instead of using a single frequency CW pump, DSF is pumped by a pulsed laser~\cite{Fiorentino02,Rarity05,Fan05a}.
In this case, the average detected single count rates (per pulse), produced by SFWM in signal and idler channels can be written as~\cite{Chen05,Li08a-opex}
\begin{eqnarray}
R_{sF}=A_s\eta _{d1}\eta _{ts}(\gamma P_pL)^2\frac{\sigma _s}{\sigma _p}\cr
R_{iF}=A_i\eta _{d1}\eta _{ts}(\gamma P_pL)^2\frac{\sigma _i}{\sigma _p}
\end{eqnarray}
where $A_s$ and $A_i$ are coefficients associated with the experimental details, $P_p$ is the peak pump power, $L$ and $\gamma$ are the length and nonlinear coefficient of the optical fiber, respectively; $\sigma _p$, $\sigma _s$ and $\sigma _i$ are the spectral bandwidth of pump, signal and idler photons determined by their corresponding filters, respectively. Thus, the coincidence rate of the photon pairs is given by
\begin{eqnarray}
C_c=\xi _i\eta _{d2}\eta _{ti}R_{sF}=\xi _s\eta _{d1}\eta _{ts}R_{iF},
\end{eqnarray}
with
\begin{eqnarray}
\xi _{s(i)}=\frac{\int d\Omega _{s(i)}f(\Omega _{s(i)})S_{s(i)}}{\int
d\Omega _{s(i)}S_{s(i)}}
\label{collection-pulse}
\end{eqnarray}
refers to the collection efficiency of the photon pairs for the heralding photons detected in idler (signal) channel, where $\Omega _{s(i)}$ is related to $\omega _{s(i)}$ and the central frequency of filter in signal (idler) channel $\omega _{s(i)0}$ by $\Omega _{s(i)}=\omega _{s(i)}-\omega _{s(i)0}$, and $f(\Omega _{s(i)})$ is a function describing the filter placed in signal (idler) channel. $S_{s(i)}=\int d\Omega _{i(s)}f(\Omega _{i(s)})\left| F(\omega _s,\omega
_i)\right| ^2$ in Eq. (\ref{collection-pulse}) is the conditional spectrum, describing the individual signal (idler) photon wave packet for the heralding idler (signal) photons with a spectrum shaped by the filter function $f(\Omega _{i(s)})$. According to the Ref. \cite{Li08a-opex}, for a Gaussian shaped pump pulse, the spectrum function can be expressed as
\begin{eqnarray}
F(\omega _s,\omega _i)&=&\int\limits_{-L}^0dz\frac{\exp \left\{ -i\Delta
kz-2i\gamma P_pz\right\} }{\sqrt{1-ik^{^{\prime \prime }}\sigma
_p^2z-\frac i2k^{^{\prime \prime \prime }}(\Omega _s+\Omega _i)z\sigma _p^2}}\cr
&\times&\exp \left\{ -\frac{(\Omega _s+\Omega _i)^2}{4\sigma _p^2}\right\} ,
\end{eqnarray}
where $\Delta k$ is the wave vector mismatch; the terms $k^{^{\prime \prime }}$ and  $k^{^{\prime \prime \prime }}$ are the second and third order dispersion at the central frequency of pump, respectively.
Because of
the broadband nature of pulsed pump and bandwidth of SFWM in DSF, the coefficient $\xi _{s(i)}$ is generally less than 1 unless the bandwidth of the product $f(\Omega _{s(i)})S_{s(i)}$ is the same as that of $S_{s(i)}$, which means the bandwidth of $S_{s(i)}$ is much smaller than that of the signal (idler) photons described by $\sigma _{s(i)}$.
Under this condition, the QEs of SPDs can be expressed as
\begin{eqnarray}
\eta _{d2}=\frac{C_c}{\xi _iR_{sF}\eta _{ti}},~~\eta _{d1}=\frac{C_c}{\xi
_sR_{iF}\eta _{ts}}
\label{effi-pulse}
\end{eqnarray}
Eq. (\ref{effi-pulse}) shows that a reliable deduction of the QEs requires the accurate estimation of both the collection efficiency $\xi_{s(i)}$ and the transmission efficiency $\eta _{ts(i)}$ in signal (or idler) channel. The measurement of $\eta _{ts(i)}$ is straightforward, but the evaluation of $\xi_{s(i)}$ is not trivial.

Eq. (\ref{collection-pulse}) indicates that the function $S_s$ is an important parameter for estimating $\xi _s$. In general, $S_s$ can not be analytically solved out, because of the complicated dependence of spectrum function $F(\omega _s,\omega _i)$. However, it is possible to deduce the expression of $S_s$ via a experimental measurement described in Ref. \cite{Li08a-opex}. In the experiment, the pump and idler photons with Gaussian shaped spectra are fixed. At a certain power level, after measuring the true coincidences of photon pairs as the central wavelength of the Gaussian shaped filter in signal channel, $\lambda _{s0}\prime$, is scanned, then plot the true coincidences as a function of $\lambda _{s0}\prime$ and fit the measured data with a function $S_{scan}$. It is straightforward to figure out that $S_{scan}$ is associated with $S_s$ and the filter in signal channel $h(\omega _{s}-\omega _{s0}\prime)$ through the relation $S_{scan}=\int d\omega _{s}h(\omega _{s}-\omega _{s0}\prime)S_{s}$, and thus $S_s$ can be deduced accordingly. In the calculation, the filter in signal channel is described by
\begin{eqnarray}
h(\omega _{s}-\omega _{s0}\prime)\propto \exp (-\frac{(\omega _s-\omega _{s0}\prime)^2}{%
\sigma _s^2}),
\label{scanning-1}
\end{eqnarray}
where $\omega _{s0}\prime$ refers to the central frequency of the filter in the signal channel, and the fitting function of the measured true coincidence is
\begin{eqnarray}
S_{scan}\propto \exp (-\frac{(\omega _{s0}\prime-\omega _{s0})^2}{(\sigma _0^{\prime })^2}),
\label{scanning-2}
\end{eqnarray}
where $\sigma _0^{\prime }$ is the bandwidth of $S_{scan}$. In this situation, we have
\begin{eqnarray}
S_{s}\propto \exp (-\frac{\Omega _s^2}{(\sigma _0)^2}),
\label{scanning-3}
\end{eqnarray}
where $\sigma _0= \sqrt{{\sigma _0^{\prime }}^2-\sigma _s^2}$.

It is worth noting that $S_s$ can have an analytical expression and $\xi_{s(i)}$ can be estimated by a theoretical calculation when some specific conditions are satisfied. For example, if the phase matching condition is perfectly satisfied, and the bandwidth of the pump pulse is so narrow that the pulse broadening due to self-phase modulation (SPM) and dispersion is negligibly small, the spectrum function has the simplified form
$F(\omega _s,\omega _i)= L \exp ( -\frac{(\Omega _s+\Omega _i)^2}{4\sigma _p^2})$.
Therefore, we have the conditional spectrum
\begin{eqnarray}
S_s\propto \exp (-\frac{\Omega _s^2}{2\sigma _p^2+\sigma _i^2}).
\label{analitical-Ss}
\end{eqnarray}
In this situation, substituting the filter function in signal channel $f(\Omega _{s})$  into Eq. (\ref{collection-pulse}), the collection efficiency can be written as
\begin{eqnarray}
\xi _s=\frac{\int d\Omega _{s}f(\Omega _{s})\exp (-\frac{\Omega _s^2}{2\sigma _p^2+\sigma _i^2})}{\int
d\Omega _{s}\exp (-\frac{\Omega _s^2}{2\sigma _p^2+\sigma _i^2})}
\label{collection-appro}
\end{eqnarray}
For the function $f(\Omega _{s})$ describing a super-Gaussian and Gaussian shaped filter, respectively, we plot $\xi _s$ as a function of the ratio $\frac{\sigma _s}{\sigma _0}$, where $\sigma _0=\sqrt{\sigma _i^2+2\sigma _p^2}$ refers to the bandwidth of $S_s$. As shown in Fig. 1, one sees that when the two kinds of filters have the same bandwidth $\sigma _s$, super-Gaussian shaped filter corresponds to a higher collection efficiency. Moreover, the point $\xi _s=99 \%$ corresponds to $\frac{\sigma _s}{\sigma _0}=2.3$ and $\frac{\sigma _s}{\sigma _0}=7$ for $f(\Omega _{s})$ with a super-Gaussian and Gaussian shaped spectra, respectively. Therefore, it is easier to get a higher collection efficiency by using commercially available super-Gaussian shaped WDM filters in heralded signal channel~\cite{Li06}.

Apart from estimating the parameter $\xi _s$, according to Eq. (\ref{effi-pulse}), a correct estimation of the parameter $R_{i(s)F}$, denoting the photon production rate in idler (signal) band via SFWM, is also important for the photon pair based QE measurement. Moreover, for an efficient SFWM process, the detuning between signal (idler) and pump photons is usually less than a few THZ for the pump with the central wavelength close to the zero dispersion wavelength of DSF, so the detuning is often chosen to be smaller to suppress Raman scattering~\cite{Li04}. This may cause photons in the signal and idler bands originated from SPM of pump, $R_{SPMs(i)}$, also propagate through the signal and idler channels~\cite{Li06, Li05c}. Therefore, besides reliably subtracting Raman scattering from the total scattered photons in idler (signal) channel, photons generated via SPM should also be reliably excluded.

To obtain $R_{i(s)F}$ with high accuracy, the detuning of photon pairs is preferred to be large enough so that $R_{SPMi(s)}$ is negligibly small. In this case, the total measured rate of photons in heralding idler (signal) band can be fitted with the equation
\begin{eqnarray}
N_{T}=s_1P_{ave} + s_2P_{ave}^2,
\label{fitting}
\end{eqnarray}
where $P_{ave}$ is the average power of the pulsed pump, $s_1$ and $s_2$ are the linear and quadratic scattering coefficients, which respectively determine the strengths of Raman scattering and SFWM, $R_{Ri(s)}$ and $R_{i(s)F}$. In the experiment of QE measurement, to minimize the estimation uncertainty of $R_{i(s)F}$, instead of extracting its value from the fitting parameters $s_2P_{ave}^2$ in Eq. (\ref{fitting}), we first separately measure the total scattered photons in idler (signal) channel $N_{T}\prime=R_{Ri(s)}=s_1P_{ave}$ by adjusting the central wavelength of pump in the normal dispersion regime~\cite{Li05c}, and then subtract the Raman scattering $R_{Ri(s)}$ from $N_{T}$.

\section{Photon counting detection system}

The signal and idler photons are detected by InGaAs/InP avalanche photodiodes based SPDs (id200 and PLI-AGD-SC-Rx) operated in the gated Geiger mode, respectively. The gate pulses arrive at a rate of about $1.29$\,MHz, which is $1/32$ of the repetition rate of the pump pulses, and the dead time of the gate
is set to be 10 $\mu$s. The pulse widths of the gate for the two APDs are 2.5 ns and 1 ns, respectively. The timing of the gate pulses are adjusted by a digital delay generator to coincide with the arrival of signal and idler photons which are generated in DSF by a pulsed pump. The electrical signals produced by the SPDs in response to the incoming photons (and dark events), reshaped into 100-ns wide
TTL pulses, are then
acquired by a computer-controlled analog-to-digital (A/D)
board (National Instrument, PCI-6251). Thus, single counts in
both the signal and idler channels and coincidences acquired from
different time bins can be determined because the A/D card records
all counting events.

The relative timing drift between the electrical gate and the arrival of signal and idler photons due to the imperfectness of the electrical circuits may cause the fluctuation in single counts and coincidences. In our system, the fluctuation  accumulated in one hour is less than $3\%$ for both detectors. For all the photon counting measurements presented in the paper, the integration time at each data point is less than 10 minutes, this fluctuation is almost unobservable. Therefore, we estimate the uncertainty of the measured QE contributed by the fluctuations is less than $0.5\%$

Under the gate rate of $1.29$\,MHz, with a dead time of 10 $\mu$s, the dark-count probabilities of the two SPDs (id200 and PLI-AGD-SC-Rx) are $1.7
\times 10^{-5}$/pulse and $3 \times 10^{-5}$/pulse, respectively; the after-pulse probabilities for both SPDs,  directly estimated from the raw data in dark conditions with the method described
in reference~\cite{Voss04}, are less than 0.005. In the photon counting measurements, the dark counts are subtracted. The after-pulse effect will bring false triggers in the QE measurement. According to Eq. (\ref{effi-pulse}), the inaccuracy of the measured efficiency of SPD originated from after-pulse effect will be less than $0.5\%$. As for the dead time of SPD, which results in a reduction of the sampling rate and does not affect the evaluation of the trigger rate and coincidence rate in Eq. (\ref{effi-pulse}), will not contribute to the measurement uncertainty.

\section{Experiment}

After understanding the experimental principle and the basic characteristics of the photon counting system, we perform a series of experiments to realize the calibration of photon-counting detection efficiency.

The experimental setup is shown in Fig. 2. Signal and idler photons are produced
by pumping 300\,m DSF with laser pulses. The zero dispersion wavelength of DSF is $\lambda
_0=1544\pm 2$ nm at room temperature. The pump pulses with a pulse width of about $10$ pico-second are spectrally
carved out from a mode-locked femto-second fiber laser with a
repetition rate of about 41 MHz. To achieve the required power, the pump
pulses are amplified by an erbium-doped fiber amplifier.
The pump pulses are further cleaned up with a band-pass filter
F$_1$, which is realized by cascading a tunable filters (Newport/TBF-1550-1.0) with one channel of array-waveguide-grating (AWG), and the full width at half-maximum (FWHM) of the Gaussian shaped pump pulse is 0.3 nm. Passing through a fiber
polarization controller (FPC1) and a polarization beam splitter
(PBS1) to ensure the polarization and pump power adjustment, a 90/10 fiber coupler is used
to split $10\%$ of the pump for power monitor and for estimating the transmission efficiency of signal and idler photons.

Signal and idler photons co-polarized with the pump are selected by adjusting the fiber polarization controller (FPC2). To reliably detect the signal and idler photons, an
isolation between the pump and signal/idler photons in excess of
100\,dB is required, because of the low efficiency of SFWM in DSF. In addition, detecting signal and idler photons with
different bandwidths is also necessary for studying the correlation dependence of photon pairs .
We achieve these by passing the photon pairs through a filter
ensemble F$_2$, whose FWHM can be adjusted from 0.15 nm to 2.2 nm. F$_2$ is realized either by using double grating filters (DGFs) or WDM filters, or by cascading WDMs with a DGF, a fiber Bragg-grating (FBG) filter, or one channel of AWG. The overall transmission efficiencies for the signal and idler photons are determined by the splicing loss between DSF and the standard single mode fiber, and by the transmission efficiencies of FPC2, polarization beam splitter (PBS2) and F$_2$, while the transmission loss in 300\,m DSF is only about 0.1 dB.

The signal and idler photons are detected by the photon counting system described in the previous section. In our experiments, SPD (PLI-AGD-SC-Rx, labeled as SPD$_T$), serving as a trigger detector, is used to record the heralding idler photons, and SPD (id200, labeled as SPD$_{UT}$) with the quantum efficiency $\eta _{UT}$ under test, is used to record the heralded signal photons. According to Eq. (\ref{effi-pulse}), $\eta _{UT}$ can be written as
\begin{eqnarray}
\eta_{UT}=\eta_{d1}=\frac{C_c}{\xi _sR_{iF}\eta _{ts}}.
\label{qe-ut}
\end{eqnarray}

To extract photons produced via SFWM from the measured total counts in idler channel, $R_{iF}$, we first measure Raman scattering when the detuning between the idler and pump photons is $\Omega=0.83$ THz. At such a detuning, for the pulsed pump with a FWHM of 0.3 nm and with the power within our concerned level, the photons in signal and idler bands originated from SPM are negligible~\cite{Ma10}. In the experiment, the central wavelength of pump is adjusted to 1530 nm, so that phase matching for SFWM is not satisfied, and only the photons originated from Raman scattering are measured. To estimate the transmission efficiency $\eta _{ti}$, we first block the pump and SPDs, then tune the central wavelength of Santec laser (TSL-210V) to that of F$_2$ in idler channel, and launch its output from the 10 $\%$ port. Comparing the input power of DSF and the output power of F$_2$, the value of $\eta _{ti}$ is determined. While during the photon counting measurement, the Santec laser is turned off. The Raman scattering measurement is made by using F$_2$ with various bandwidth in idler channel. In each case, we record the single counts in idler channel by changing the pump power. Then we fit the measured data with the function
\begin{eqnarray}
N_{T}\prime / \eta _{ti}=R_{Ri}/\eta_{ti}=(s_1 / \eta _{ti})P_{ave}=s_1\prime P_{ave},
\label{normalized-s1}
\end{eqnarray}
where $s_1\prime $ is the normalized coefficient of $s_1$, and the units of $R_{Ri}$ and $P_{ave}$ are $10^{-3}$ photons/pulse and mW, respectively. Table 1 shows the coefficient $s_1\prime $ for F$_2$ in idler channel with the bandwidth of 0.15, 0.46, 0.69, 1.02, and 1.1 nm, respectively.

In the first quantum efficiency measurement, we perform photon counting measurements to obtain the parameters $C_c$ and $R_{iF}$, and $\xi _s$ is evaluated by performing the experiment similar to Ref.~\cite{Li08a-opex}. In the experiment, the central wavelengths of pump, signal and idler photons are $\lambda _p=
1544$\,nm, $\lambda _s=
1550.7$\,nm and $\lambda _i=
1537.4$\,nm, respectively; and the FWHM of Gaussian shaped signal and idler photons are 0.60 and 1.02 nm, respectively. We first record the single counts and coincidences of signal and idler photons as a function of the pump power. The true coincidence $C_c$ is obtained by subtracting the accidental coincidences, calculated from the single counts in signal and idler channels, from the measured coincidences produced by the same pump pulse. Again, the transmission efficiency $\eta _{ts(i)}$ is estimated by blocking the pump and SPDs, then launching Santec laser at the corresponding wavelength from the 10 $\%$ port. According the the acquired data, we plot the single counts in idler band $N_{T}$ versus pump power, and fit the data with the function $N_{T}=\eta _{ti}s_1\prime P_{ave} + s_2P_{ave}^2$, where $R_{iF}=s_2P_{ave}^2$, as shown in Fig. 3(a). The inset of Fig. 3(a) shows the true coincidence $C_c$ as a function of $ \eta _{ts} R_{iF}$. To reduce the uncertainty of $\eta_{UT}$ and to ensure the photon pair production rate is small enough to reliably deduce $\eta_{UT}$, we take the count rate $C_c$ and $R_{iF}\eta _{ts}$ obtained at the average pump power of about 0.18 mW, wherein the corresponding production rate is about $0.01$ pairs/pulse.

To estimate $\xi _s$, we then perform a set of measurements by fixing the spectra of the pump and idler photons and scanning the central wavelength of the Gaussian shaped filter in signal channel $\lambda_{s0}\prime$, whose spectrum is shown in the inset of Fig. 3(b). At the average pump power of 0.18 mW, we record single counts and coincidences at each $\lambda _{s0}\prime$. To normalize the deduced true coincidences, we also estimate the transmission efficiency $\eta _{ts}$ in signal channel after making each measurement. Figure 3(b) plot $C_c$ versus $\lambda _{s0}\prime$, and the solid curve is the fitting of Gaussian functions with FWHM of $1.22\pm 0.06$ nm. According to Eq. (\ref{collection-pulse}) and Eq. (\ref{scanning-1})-(\ref{scanning-3}), we obtain $\xi _s=0.496 \pm 0.03$. Thus, we can deduce $\eta _{UT}=(11.7\pm 1.5)\%$.

Considering our experimental parameters satisfy the required specific conditions for deducing Eq. (\ref{analitical-Ss}), we then conduct a series photon counting measurement to investigate if collection efficiency $\xi _s$ can be theoretically calculated. In the experiment, the parameters are the same as before, except the setting of F$_2$. In the data acquisition process, we first make measurements by using the filter $F_2$, having a sixth-order super-Gaussian shaped spectrum with a FWHM of 1.15 nm in signal channel, and the FWHM of the Gaussian shaped idler spectrum is adjusted to be 0.15, 0.46, 0.69, and 1.1 nm, respectively; we then make measurements by changing the dual-band filter $F_2$ so that the FWHMs in super-Gaussian shaped signal and Gaussian shaped idler bands are 0.67 and 1.1 nm. The spectra of $F_2$ with different bandwidths are shown in Fig. 4(a)-(c). In addition, to ensure SFWM in DSF is in the low gain regime, i.e., pair production rate is smaller than 0.05 pairs/pulse, the average power of pump is controlled to be less than 0.3 mW. Under the power level, the pulse broadening due to SPM is negligibly small, which is proven by the spectra of pump in the input and output of DSF, as shown in Fig 4(d).

For each kind of setting of $F_2$, we record the single counts and coincidences of signal and idler photons as a function of the pump power, and the true coincidence $C_c$ is deduced accordingly. Also, for each set of measurements, we extract $R_{iF}$ from the measured total counts $N_{T}$ through the relation $R_{iF}=N_{T}-\eta _{ti}s_1\prime P_{ave}$, and the transmission efficiency $\eta _{ts(i)}$ is successively estimated.
Based on the measurement, we plot the true coincidence $C_c$ as a function of $ \eta _{ts} R_{iF}$, and fit the results with $C_c=\zeta  R_{iF} \eta _{ts}$, as shown in Fig. 5(a). To maintain the consistency with the theoretical simulation in Fig. 1, different set of data are labeled by $\frac{\sigma _s}{\sigma _0}$, where $\sigma _0 $ is the calculated bandwidth of the conditional spectrum $S_s$ in Eq. (\ref{analitical-Ss}).

Comparing the fitting function $C_c=\zeta  R_{iF} \eta _{ts}$ (see Fig. 5(a)) with Eq. (\ref{qe-ut}), one sees that $\zeta $ is associated with QE and the collection efficiency through the relation $\zeta =\xi _s \eta_{UT}$. We then plot the fitting coefficient $\zeta $ versus the ratio $\frac{\sigma _s}{\sigma _0}$, and make the theoretical fit $\zeta =\xi _s \eta_{UT}$ with $\xi _s$ calculated by substituting experimental parameters into Eq. (\ref{collection-appro}) and with $\eta _{UT}$ adjusted for the best fit. As shown in Fig. 5(b), each diamond corresponds to the data point $\zeta$ obtained by using the specified $F_2$ , the solid curve is the theoretical fit. It is clear that the theoretical fit agrees with the experimental result well, indicating the collection efficiency $\xi _s$ can be calculated from the experimental parameters associated with the spectra of pump, signal and idler photons.

Using the theoretically calculated results of $\xi _s$, we then measure $\eta_{UT}$ by exploiting photon pairs with various kinds of bandwidth. To reduce the uncertainty, for each set of the photon counting measurement result obtained by using $F_2$ with different bandwidth combination in signal and idler bands, we choose the data point having the maximum rate of $C_c$, whose corresponding average pump power and photon pair production rate, $P_{ave}$ and $P_{pair}$, are shown in table 2. Figure 6 presents the value of $\eta_{UT}$ obtained by using signal and idler photons with different ratio $\frac{\sigma _s}{\sigma _0}$. One sees the values of $\eta_{UT}$ deduced from different set of data agree with each other, and they are in consistent with the value $\eta _{UT}=(11.7\pm 1.5)\%$ obtained in previous experiment. Taking the average of the five data points, we have $\eta_{UT}=(12.1\pm 0.5)\%$.

\section{Discussion}

Having experimentally demonstrated that QE of the InGaAs/InP based SPD $\eta_{UT}$ can be deduced by using signal and idler photon pairs with various kinds of bandwidths, let's first analyze the measurement uncertainty and reliability. Then we will test the validity of the deduced $\eta_{UT}$. Finally, we discuss the how to improve the experimental results.

In our measurements, the standard deviations of the quantum efficiency $\eta_{UT}$ caused by fluctuations in the photon counting measurements can be written as
\begin{equation}
\delta \eta _{UT}  = \eta _{UT} \sqrt {\left( {\frac{{\delta C_c }}
{{C_c }}} \right)^2  + \left( {\frac{{\delta \xi _s }}
{{\xi _s }}} \right)^2  + \left( {\frac{{\delta R_{iF} }}
{{R_{iF} }}} \right)^2  + \left( {\frac{{\delta \eta _{ts} }}
{{\eta _{ts} }}} \right)^2 }
\label{error-eta}
\end{equation}
with
\begin{eqnarray}
    \delta R_{iF}  &=& \left\{\left( {\delta N_T } \right)^2  + \left(\eta _{ti} s_1\prime P_{ave}\right)^2 \left\{ \left( {\frac{{\delta \eta _{ti} }}
{{\eta _{ti} }}} \right)^2   \right.  \right.  \cr
&+& \left.\left. \left( {\frac{{\delta s_1 \prime}}
 {{s_1 \prime}}} \right)^2 +  \left( {\frac{{\delta P_{ave} }}
{{P_{ave} }}} \right)^2  \right\}\right\}^{1/2},
\label{error-Rif}
\end{eqnarray}
and
\begin{equation}
 \delta s_1{\prime }=s_1{\prime }\sqrt{\left( {\frac{{\delta R_{Ri}}}{{%
R_{Ri}}}}\right) ^2+\left( {\frac{{\delta \eta _{ti}}}{{\eta _{ti}}}}%
\right) ^2+{\left( {\frac{{\delta P_{ave}}}{{P_{ave}}}}\right) ^2}}.
\label{error-s1}
\end{equation}

The standard deviations of some parameters in Eq. (\ref{error-eta})-(\ref{error-s1}) can be straightforwardly estimated, as listed in table 3. $\frac{\delta N_T} {N_T} $, $\frac{\delta R_{Ri}} {R_{Ri}} $ and $\frac{\delta C_c} {C_c} $ lie on the statistical fluctuation of photon counting, $\frac{\delta \eta _{ti(s)}}{\eta _{ti(s)}} $ and $\frac{\delta P_{ave}}{P_{ave}} $ rest with the power fluctuation of lasers. Whereas the standard deviations $\frac{\delta R_{iF}} {R_{iF}}$ and $\frac{\delta \xi _s} {\xi _s} $ depend on the detailed experimental conditions: $\frac{\delta R_{iF}} {R_{iF}}$ is determined by the pump power, intensity of Raman scattering and the fluctuation of the total detected counts $N_T$; while $\frac{\delta \xi _s} {\xi _s} $ is not only associated with the uncertainty of the spectra of pump, signal and idler photons, but also depend on the ratio $\frac{\sigma _s}{\sigma _0}$. We would like to mention that the imperfectness of electrical circuits and after pulse effect of SPD also contributes to the uncertainty of $\eta _{UT}$, which are less than $0.5\%$. However, to simplify the uncertainty analysis, we neglect the effects due to their smallness.

According to the above analysis, for the data shown in Fig. 3, the measured uncertainties of $\xi _s $ and $R_{iF}$ are about $4\%$ and $12\%$, respectively. Therefore, the uncertainty of the deduced $\eta _{UT}$ is about $13.5\%$. For the results in Fig. 6, the uncertainty analysis is slightly complicated. According to the experimental parameters for each data point (see table 2), using the upper limits of the standard deviation $\frac{\delta \xi _s} {\xi _s}\approx1.5\% $ determined by the resolution of the optical spectrum analyzer, we obtain their corresponding $\frac{\delta R_{iF}} {R_{iF}}$ and $\frac{{\delta \eta _{UT} }}{{\eta _{UT} }}$, as listed in table 4. Taking the average of the five date points, we can deduce the standard deviation $\frac{{\delta \eta _{UT} }}{{\eta _{UT} }}\approx 4\%$.

It is worth pointing out that multi-photon pairs events of the signal and idler photons will affect the reliability of the deduced $\eta _{UT}$. Since the quantum efficiency calibration method proposed by Klyshko~\cite{Klyshko80} actually relies on the assumption that at most one photon pair is emitted at a time. But the assumption can be easily violated by the fiber based sources of photon pairs, because of the long interaction length of SFWM in optical fibers. Let's qualitatively analyze how the pair production rate influence the measured $\eta _{UT}$. Assuming  the temporal mode of idler photons is in a single mode, if the efficiencies of the detectors in signal and idler bands are $\eta_{s0}$ and $\eta_{i0}$, respectively, $\eta _{UT}$ can be written as
\begin{eqnarray}
\frac{{\eta _{UT}}}{{\eta_{s0}}}=\frac{1+\overline{n}}{(1+\overline{n}\eta _{ts}\eta _{s0})[1+%
\overline{n}(\eta _{ts}\eta _{s0}+\eta _{ti}\eta _{i0}-\eta _{ts}\eta
_{s0}\eta _{ti}\eta _{i0})]}
\label{multi-photons}
\end{eqnarray}
where $\overline{n}$ is the average photon number in idler channel via SFWM. Eq. (\ref{multi-photons}) indicate photon pairs with a higher production rate, corresponding to a higher average photon number in idler channel, will cause an increase of the deduced value of $\eta _{UT}$. Although under the condition $\eta _{ti}\eta _{i0},\eta _{ts}\eta _{s0}\ll 1$, which is generally satisfied in experiments, one have $\eta _{UT}\approx\eta_{s0}(1+\overline{n})$, measured $\eta _{UT}$ still can not be corrected accordingly. This is because the temporal mode of idler (signal) photons in our experiment is not a single mode~\cite{li08ol}. To reliably get rid of the influence of multi-photon pairs events, a detailed multi-mode analysis in the higher gain regime is necessary~\cite{Alibart06}.

According to the analysis of the influence of multi-photon pairs events, we can in principle choose to measure QE by reducing the pump power, so that the production rate of photon pairs is very low. However, the solution is not practical. Since in this case, to ensure a smaller statistical fluctuation of photon counting, we need to increase the integration time during the photon counting measurement. Limited by the long term stability of our photon counting system, this will increase the measurement uncertainty. In our QE measurement, the pair production rate is less than $\sim 3\%$, therefore we can say that the increase of the measured $\eta _{UT}$ due to multi-photon pairs events is less than $3\%$.

To test the validity of the  $\eta _{UT}$ deduced from the photon pairs-based method, we also measure the QE by utilizing a weak continuous wave (CW) laser, which is obtained by attenuating the CW Santec laser with its wavelength tuned to 1550 nm. Using the method described in Ref. \cite{Brodsky-opex10}, we first measure the intensity of the attenuated laser by using a power meter with the uncertainty of about $5\%$. Then we send the weak laser with an average power of $93$ nW through the attenuator with an attenuation of about $-42.6 \pm 0.3$ dB, which is constructed by cascading four 10/90 fiber couplers. In this case, the light intensity incident on SPD$_{UT}$ is $0.1\pm0.009$ photons/2.5-ns. Taking the measured efficient gate width ($0.62\pm0.02$ ns) of SPD$_{UT}$ into account, we obtain $\eta _{UT}=(11.7\pm 1.2)\%$. The results indicate that the values of $\eta _{UT}$ obtained by using the correlated photons and by using weak lasers agree with each other.

Comparing with the high accuracy quantum efficiency measurement performed by using photon pairs generated via SPDC in $\chi ^{(2)}$ crystals~\cite{Migdall07-opex}, the uncertainty of $\eta _{UT}$ measured by using the fiber based source of photon pairs is quite high. We think the accuracy of our current measurement is mostly limited by the estimated uncertainty of $\eta_{ts(i)}$ originated from the power fluctuation of the Santec laser, which is about $4\%$, because the uncertainty also transforms to other parameters, such as $R_{iF}$ and $s_1\prime$. We believe the uncertainty of $\eta_{ts(i)}$ can be reduced to less than $1\%$ by using a better calibration scheme.
In addition, the temperature of DSF also affect the accuracy of $\eta _{UT}$. According to Eq. (\ref{error-Rif}), increasing the portion of Raman scattering in the total counts $N_T$ will increase  uncertainty of $R_{iF}$. Therefore, cooling the fiber will help to suppress Raman scattering and to decrease the uncertainty of $\eta _{UT}$.

\section{Conclusion}
In conclusion, based on our investigation of the relation between the collection efficiency and correlation dependence of the photon pairs, we have performed the proof-of-principle demonstration of absolute efficiency measurement of SPD by using photon pairs generated in optical fiber. Thanks to the advantage of modal purity, the efficiency of the InGaAs/InP based SPD $\eta_{UT}$ can be deduced by using signal and idler photon pairs with various kinds of bandwidths. Whereas for the absolute calibration carried out by using the $\chi^{(2)}$ based photon pairs, the bandwidth of the heralded photons is required to be much broader than that of the heralding photons so that all the pair events of the heralding photon are caught. Moreover, the accuracy of our quantum efficiency measurement can be improved by suppressing the Raman scattering, by evaluating the transmission efficiency in signal and idler channel more precisely, and by using the multi-mode analysis in the higher gain regime to study the influence of multi-photon pairs event and to reliably make the correction. Furthermore, our study helps to have a better understanding to the spectra correlation of the fiber based photon pairs, therefore, it is also useful for developing other photon pair based quantum information technologies.

\section*{Acknowledgement}
This work was supported in part by NCET-060238, the NSF of China
(No. 10774111), Foundation for Key Project of
Ministry of Education of China (No. 107027), the Specialized Research Fund for the Doctoral Program of Higher Education of China (No. 20070056084), 111 Project B0704, and the State Key
Development Program for Basic Research of China (No. 2010CB923101)

\bibliographystyle{osajnl}

\newpage

Fig. 1. (Color online) Collection efficiency $\xi _s$ as a function of $\frac{\sigma _s}{\sigma _0}$. The solid and dashed curves are the calculated results for $f(\Omega _{s})$ described by a sixth-order super-Gaussian and a standard Gaussian functions, respectively.

Fig. 2. (Color online) A schematic of the experimental setup. EDFA, erbium-doped fiber amplifier; F1, filter; FPC, fiber polarization controller; PBS, polarization beam splitter; F$_2$, dual-band filter.

Fig. 3. (Color online) (a) Single counts in idler band $N_{T}$ as a function of pump powers. The solid curve is the fit of the polynomial $N_{T}=\eta _{ti}s_1\prime P_{ave} + s_2P_{ave}^2$, the contributions
of linear scattering $\eta _{ti}s_1\prime P_{ave}$ (dash line) and quadratic scattering (dash dot curve) $R_{iF}=s_2P_{ave}^2$
are plotted separately as well. The inset shows the true coincidence $C_c$ as a function of $ \eta _{ts} R_{iF}$. (b) True coincidences versus the central wavelength of signal photons $\lambda_{s0}\prime$ for the average pump power of 0.18 mW. The solid curve is the fit of the Gaussian function $S_{scan}\propto \exp(-\frac{(\lambda _s-1550.72)^2}{0.73^2})$. The inset is the pass-band of the filter in signal band, solid curve overlapped with data points is the fitting of Gaussian function $f(\lambda_s )=\exp (-\frac{(\lambda_s -\lambda_{s0}\prime)^2}{0.36^2})$.

Fig. 4. (Color online) Pass-band spectra of the signal, idler and pump photons. Solid curves are fits to the data. (a) Pass-bands of filters in idler band fitted with Gaussian functions $f(\lambda )=\exp (-\frac{(\lambda -1537.4)^2}{0.09^2})$ and $f(\lambda )=\exp (-\frac{(\lambda -1537.4)^2}{0.46^2})$, where FWHMs are 0.15 and 0.69 nm, respectively. (b) Pass-bands of filters in idler band fitted with Gaussian functions $f(\lambda )=\exp (-\frac{(\lambda -1537.4)^2}{0.27^2})$ and $f(\lambda )=\exp (-\frac{(\lambda -1537.4)^2}{0.66^2})$, where FWHMs are 0.46 and 1.1 nm, respectively. (c) Pass-bands of filters in signal band fitted with super-Gaussian functions $f(\lambda )=\exp (-\frac{(\lambda -1550.7)^6}{0.36^6})$ and $f(\lambda )=\exp (-\frac{(\lambda -1550.7)^6}{0.6^6})$, where FWHMs are 0.67 and 1.15 nm, respectively. (d) Spectra of pump pulses with an average power of about $0.3$\,mW. Triangles and diamonds representing the spectra of pump in the input and output of DSF are overlapped. The fitting function $f(\lambda )=\exp (-\frac{(\lambda -1544)^2}{0.18^2})$ are over lapped with the data points.

Fig. 5. (Color online) (a) True coincidence as a function of $ \eta _{ts} R_{iF}$ for signal and idler photons with different ratio $\frac{\sigma _s}{\sigma _0}$. Solid lines are fits to the function $C_c=\zeta  R_{iF} \eta _{ts}$ with $\zeta $ as the fitting parameter, whose value is 0.058, 0.086, 0.106, 0.115 and 0.123 for $\frac{\sigma _s}{\sigma _0}$ equals to 0.48, 0.85, 1.27, 1.68 and 2.32, respectively. The inset is the enlargement of some data points which are not clear in the main plot due to limited space. (b) The parameter $\zeta$ as a function of $\frac{\sigma _s}{\sigma _0}$. The solid curve is the theoretical fit $\zeta =\xi _s \eta_{UT}$ with $\eta _{UT}=12.26\%$ as the fitting parameter.

Fig. 6. The quantum efficiency $\eta_{UT}$ obtained by using signal and idler photons with different ratio $\frac{\sigma _s}{\sigma _0}$.

\begin{figure}[htb]
\centering
\includegraphics[width=8cm]{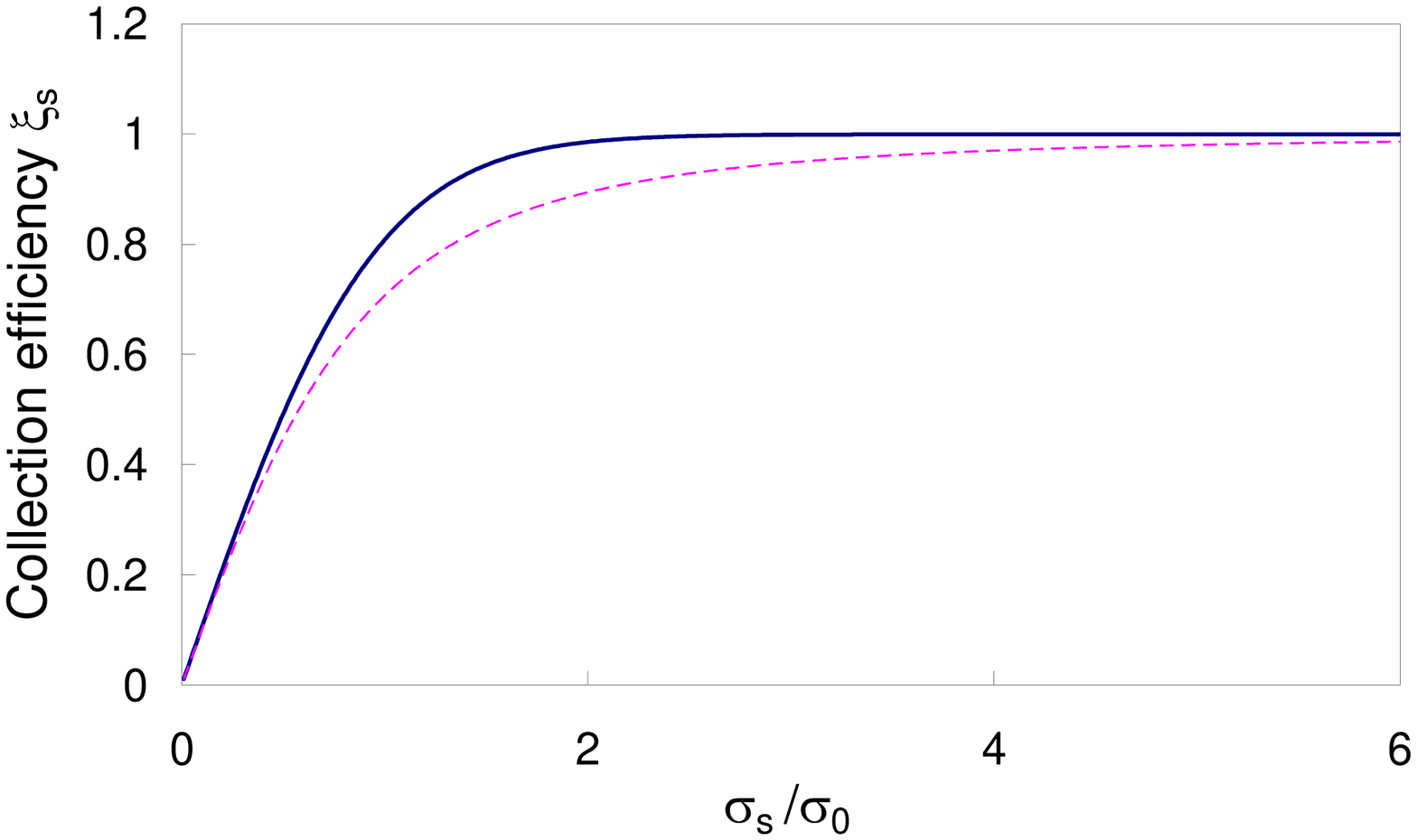}
\caption{
}
\end{figure}

\begin{figure}[htb]
\centering
\includegraphics[width=8cm]{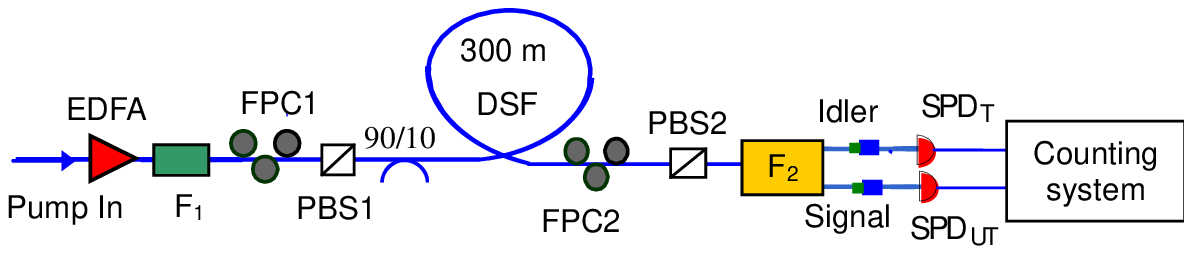}
\caption{ }
\end{figure}

\begin{figure}[htb]
\centering
\includegraphics[width=8cm]{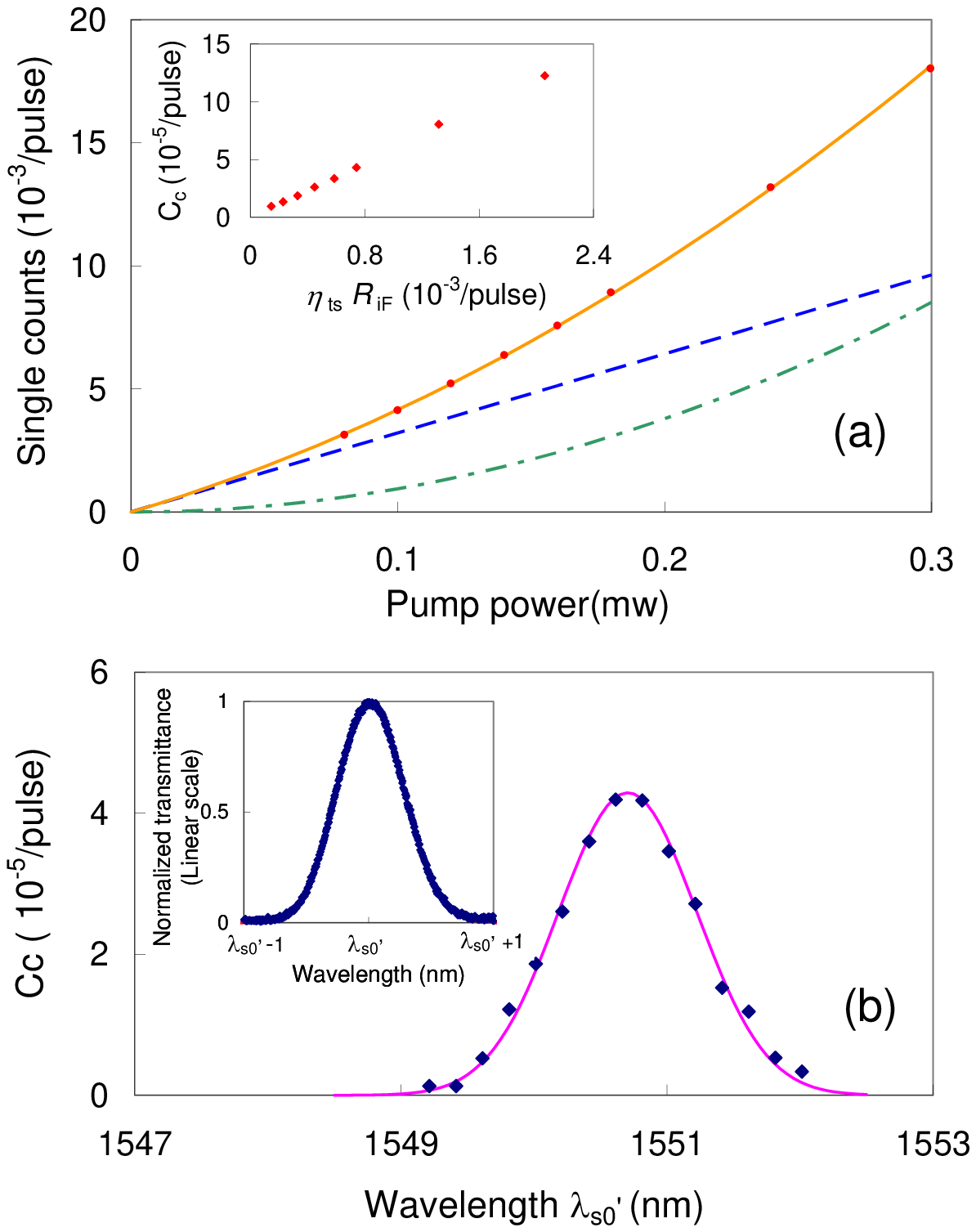}
\caption{
}
\label{}
\end{figure}

\begin{figure}[htb]
\centering
\includegraphics[width=8cm]{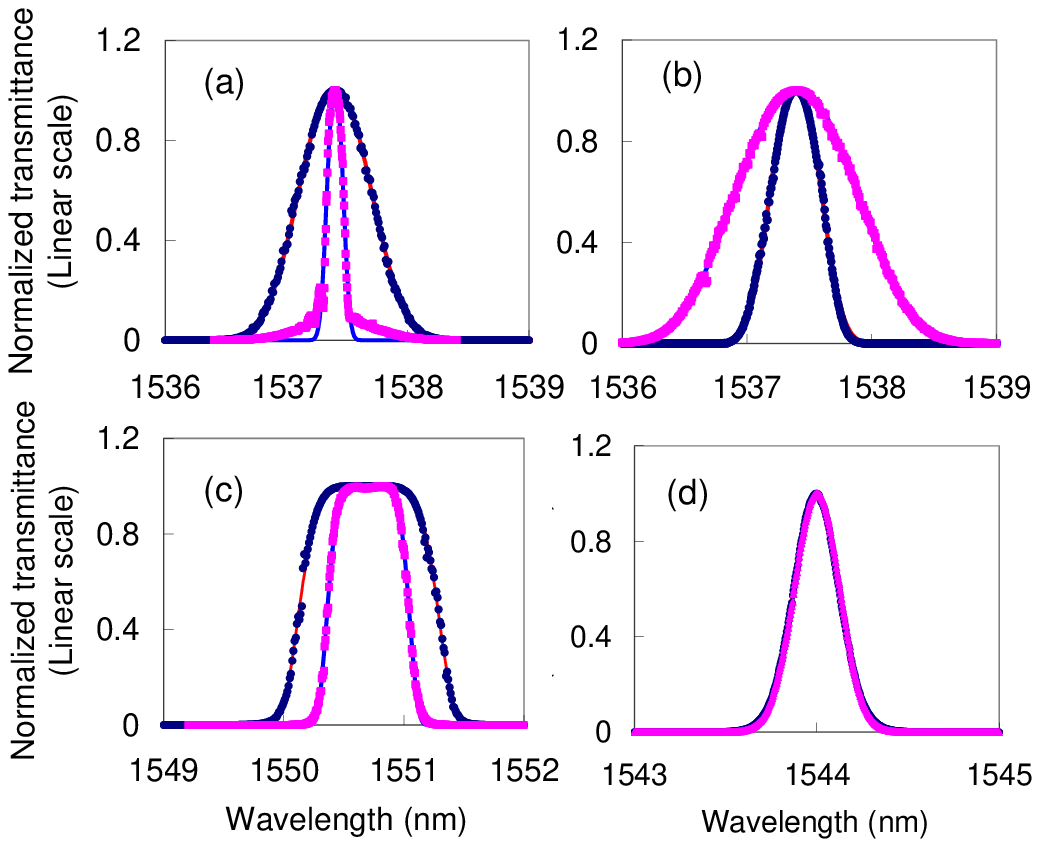}
\caption{
}
\end{figure}

\begin{figure}[htb]
\centering
\includegraphics[width=8cm]{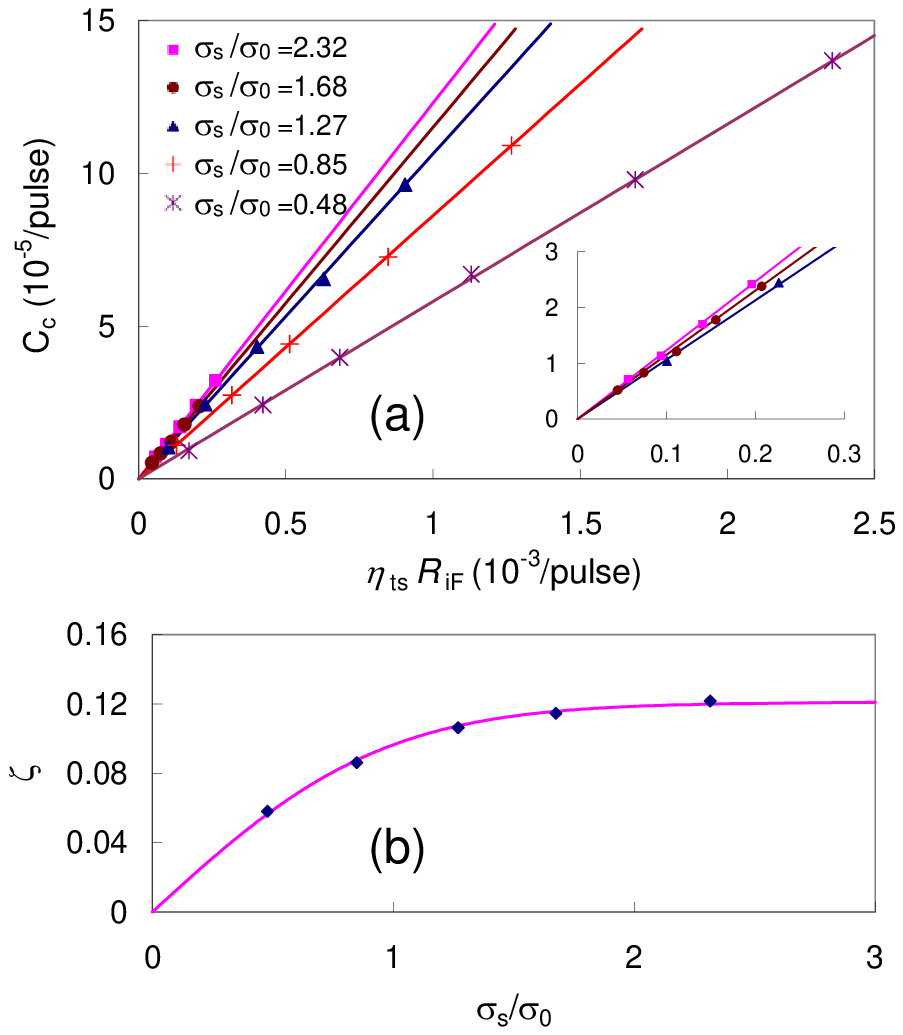}
\caption{}
\end{figure}

\begin{figure}[htb]
\centering
\includegraphics[width=8cm]{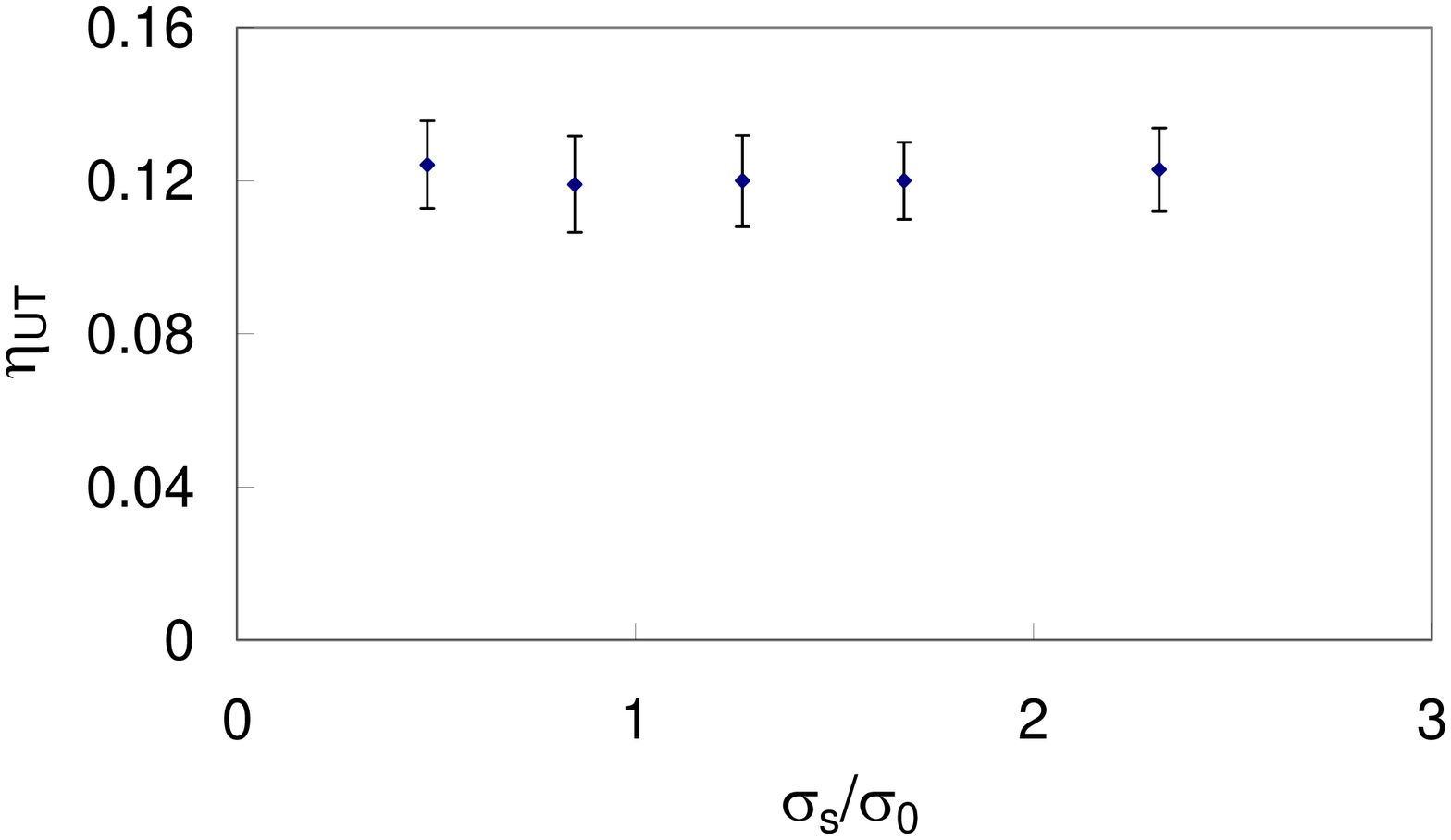}
\caption{}
\end{figure}

\begin{table}[H]
\centering\caption{Coefficient $s_1\prime$ in idler channel with different bandwidths. }
 \begin{tabular}{ @{\hspace{0.35cm}}c@{\hspace{0.35cm}}@{\hspace{0.35cm}}c@{\hspace{0.35cm}}@{\hspace{0.35cm}}c@{\hspace{0.35cm}}}
    \hline
  Bandwidth (nm) &  $s_1\prime$    \\
\hline
    0.15$\pm$0.01 & 11.93$\pm$0.54\\
    0.46$\pm$0.01 & 36.5$\pm$1.6  \\
    0.69$\pm$0.01 & 55.99$\pm$2.5  \\
    1.02$\pm$0.01 &  83.5$\pm$3.7 \\
    1.1$\pm$0.01 &  91.7$\pm$4.1 \\
  \hline
  \end{tabular}
\end{table}

\begin{table}[H]
\centering\caption{Experimental parameters for the QE }{ measurement with different F$_2$.}
\begin{tabular}{ c@{\hspace{0.2cm}}*{3}{@{\hspace{0.3cm}}c@{\hspace{0.2cm}}} }
  \hline
   $\frac{\sigma_{s}}{\sigma_{0}}$ & $P_{ave}$ (mw)  & $P_{pair}$  (pairs/pulse) \\
\hline
     2.32 &  0.3 &  $\sim 0.01$\\
       1.68 & 0.3 & $\sim 0.03$ \\
       1.27 & 0.25 & $\sim 0.03$\\
       0.85 & 0.23 & $\sim 0.03$ \\
       0.48& 0.27 & $\sim 0.03$ \\
       \hline
  \end{tabular}
\end{table}

\begin{table}[H]
\centering\caption{Standard deviations of some parameters}
 \begin{tabular}{ c@{\hspace{0.02cm}}*{5}{@{\hspace{0.2cm}}c@{\hspace{0.2cm}}} }
\hline
   & $\frac{\delta \eta _{ti(s)}}{\eta _{ti(s)}} $
   & $\frac{\delta P_{ave}}{P_{ave}} $
   & $\frac{\delta N_T} {N_T} $
   & $\frac{\delta R_{Ri}} {R_{Ri}} $
   & $\frac{\delta C_c} {C_c} $
    \\
\hline
     Standard \\Deviation($\%$) & 4 & 2 & $<$0.1 & $<$0.1 & $<$1 \\
     \hline
  \end{tabular}
\end{table}

\begin{table}[H]
\centering\caption{The standard deviations of $R_{iF}$ and $\eta _{UT}$.}
 \begin{tabular}{ c@{\hspace{0.2cm}}*{3}{@{\hspace{0.3cm}}c@{\hspace{0.2cm}}} }
  \hline
   $\frac{\sigma_{s}}{\sigma_{0}}$ & $\frac{\delta R_{iF}} {R_{iF}}(\%) $ & $\frac{\delta \eta_{UT}} {\eta_{UT}} (\%)$ \\
\hline
     2.32 &  7.85 &  8.84\\
       1.68 & 7.37 & 8.43 \\
       1.27 & 8.94 &  9.85\\
       0.85 & 9.75 & 10.58 \\
       0.48& 8.27 & 9.23 \\
       \hline
  \end{tabular}
\end{table}

\end{document}